\documentclass[journal=nalefd,manuscript=article]{achemso}

\usepackage[version=3]{mhchem} 

\author{Ryan Soklaski}
\affiliation{Department of Physics, Washington University in St.
Louis, St. Louis, MO 63136, USA}

\author{Yufeng Liang}
\affiliation{Department of Physics, Washington University in St.
Louis, St. Louis, MO 63136, USA}

\author{Changjian Zhang}
\affiliation{School of Electrical and Computer Engineering,
Cornell University, Ithaca, New York 14853, USA}

\author{Haining Wang}
\affiliation{School of Electrical and Computer Engineering,
Cornell University, Ithaca, New York 14853, USA}

\author{Farhan Rana}
\affiliation{School of Electrical and Computer Engineering,
Cornell University, Ithaca, New York 14853, USA}

\author{Li Yang}
\email{lyang@physics.wustl.edu}
\affiliation{Department of Physics, Washington University in St.
Louis, St. Louis, MO 63136, USA}

\title[An \textsf{achemso} demo]
  {Temperature Renormalization of Optical Spectra of Monolayer MoS$_2$}

\abbreviations{}

\begin{document}
\begin{abstract}
Newly measured optical absorption and photoluminescence spectra
reveal substantial frequency shifts of both exciton and trion
peaks as monolayer MoS$_2$ is cooled from 363 K to 4 K.
First-principles simulations using the GW-Bethe-Salpeter Equation
approach satisfactorily reproduce these frequency shifts by
incorporating many-electron interactions and the thermal expansion
of the in-plane lattice constant. Studying these temperature
effects in monolayer MoS$_2$ is crucial for rectifying the results
of room-temperature experiments with the previous predictions of
zero-temperature-limit simulations. Moreover, we estimate that the
thermal expansion coefficient of monolayer MoS$_2$ is around 25
$\%$ less than that of bulk counterpart by tracking the frequency
shifts of the exciton or trion peaks in optical spectra. This may
serve as a convenient way to estimate thermal expansion
coefficients of general two-dimensional chalcogenides.
\end{abstract}


Recently, much focus from the physics, chemistry, and material
science communities has been directed on the unique electronic,
magnetic, and optical properties of two-dimensional (2D)
molybdenum and tungsten chalcogenides, such as MoS$_2$
\cite{2010wang,2010mak,2011kis,2012feng,2012cui,2012heinz}.
Essential properties of monolayer MoS$_2$ and similar materials
have been heavily scrutinized by experimental and theoretical
studies. In particular, enhanced many-electron effects, including
excitons and trions, have been identified by first-principles
simulations, and their results have been widely applied to compare
and explain experimental measurements \cite{2013heinz,
exciton-1,exciton-2,exciton-3,exciton-4}. However, the comparison
of different simulations with experimental data has proven to be
controversial.

Despite the rigor and intensity with which monolayer MoS$_2$ has
been studied, a subtle yet important oversight persists: current
experimental measurements of the optical absorption spectrum of
the system are made at the room temperature
\cite{2010wang,2010mak}, while available first-principle
calculations simulate monolayer MoS$_2$ at the 0-K limit. The
marked resemblance between these experimentally and theoretically
derived spectra \cite{exciton-1,exciton-2,exciton-3,exciton-4} has
made their comparison, without regard to the temperature
discrepancy, pervasive, although temperature effects have been
known to impact band gaps and optical spectra of bulk
semiconductors. Diamond and silicon crystals are ready examples,
in which many-electron effects and electron-phonon (\emph{e-ph})
interactions induce band-gap renormalization according to
temperature variations \cite{1992cardona, 2010cohen}. In this
sense, considering temperature effects may be crucial for
rectifying recent discrepancies between theory and experiments.

Beyond affecting electronic structures themselves,
temperature-related many-electron excitations provide precious
information for detecting thermal properties of atomistic
structures. Previous studies show that the variation of the
lattice constant of monolayer MoS$_2$ will induce substantial
changes in its absorption spectra \cite{2012liu}. Therefore, by
tracking optical excitations versus temperature, we may be able to
infer the lattice variation with changing temperature, and obtain
thermal expansion coefficient (TEC), which is a fundamental
character of 2D structures but challenging to be measured
directly. For instance, the TEC of graphene and their unique
thermal behaviors have been intensively studied
\cite{2005marzari,2009lau,2011son}.

In this Letter, experimentally measured optical absorption spectra
and photoluminescence (PL) measurements taken from 4 K to 363 K
are presented alongside complementary first-principle results
about monolayer MoS$_2$. The measurements reveal that temperature
changes significantly impact optical excitations, including the
positions of exciton and trion peaks, and induce frequency shifts
of up to 50 meV in the absorption peak positions. Simulations
rooted in the GW-Bethe Salpeter equation (BSE) not only provide an
exceptional agreement with the absolute peak positions measured
from experiments at the low-temperature limit but also indicate
that the frequency shift can largely be attributed to the thermal
expansion of the in-plane lattice constant. Accordingly, we
estimate the TEC of monolayer MoS$_2$ from the shift of optical
peaks and conclude that it is around 25$\%$ smaller than that of
bulk MoS$_2$. This provides a convenient way of estimating the TEC
of 2D semiconductors.

Absorption measurements are performed on monolayer MoS$_{2}$ at
different temperatures under a microscope with 100X objective in
transmission configuration. The samples are prepared on quartz
substrates by mechanical exfoliation, and the typical size of an
exfoliated sample is $\sim200\mu m ^{2}$. Monolayer MoS$_2$
samples are identified under optical microscope and then confirmed
by Raman spectroscopy \cite{lee10}. The samples are annealed in
vacuum at 363~K for 8 hours. The samples are then placed in a
helium flow cryostat and measurements are taken at temperatures
between 5 K and 363 K. A broadband halogen light source is used to
illuminate the samples. The transmitted light is collected using a
confocal microscope setup and is dispersed by a monochromator with
a resolution of $\sim0.2$ nm. Absorption spectra are obtained by
subtracting the relative transmission spectra from unity.
Temperature-dependent PL of monolayer MoS$_2$ is collected and
analyzed using the same setup. The laser intensity is kept at
values less than $\sim200\mu W/\mu m^2$ to prevent damage to the
samples.

\begin{figure}
\includegraphics*[scale=0.60]{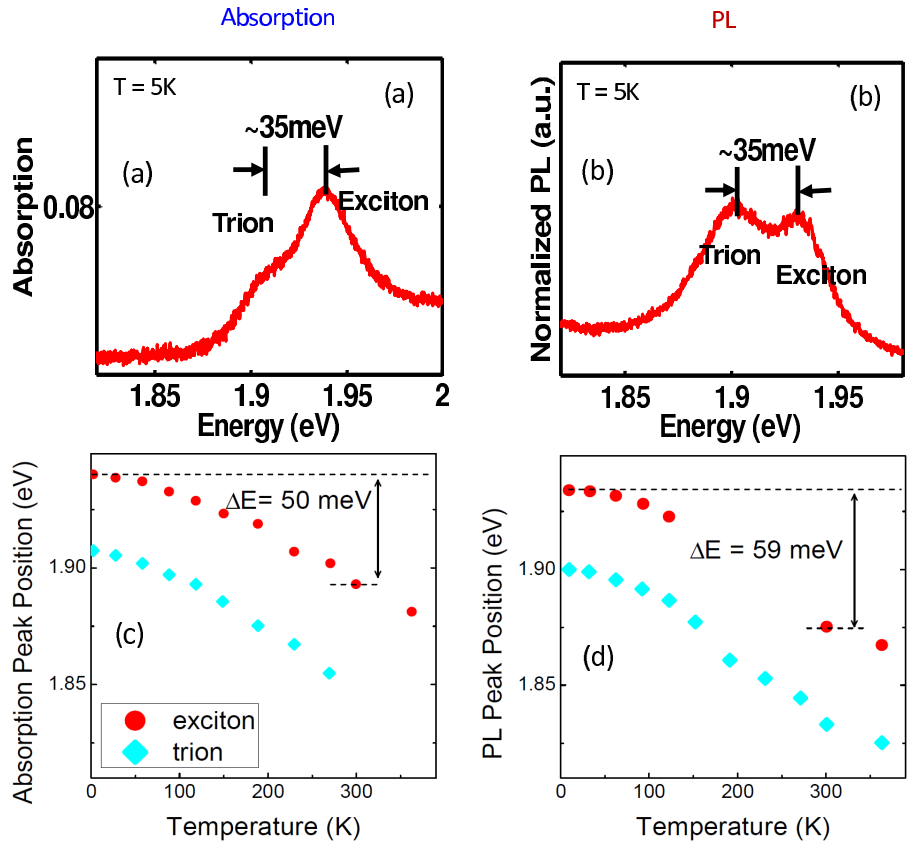}
\caption{(Color online) (a) The optical absorption spectrum of
monolayer MoS$_2$ at 4 K. (b) The PL spectrum of monolayer MoS$_2$
at 4 K. (c) The optical absorption peak positions of excitons and
trions vs temperature. (d) the PL peak positions of excitons and
trions vs temperature. } \label{exp}
\end{figure}

The measured optical absorption spectrum of monolayer MoS$_2$ at 4
K is presented in Fig.~\ref{exp} (a). The featured prominent peak
and lower-energy peak are widely attributed to exciton and trion
states, respectively \cite{2013heinz}. It is important to note
that these peak positions, at 4 K, reside systematically at higher
energies than do the peaks found in experimental data that is
collected at room temperature \cite{2010wang,2010mak}. Additional
measurements of the optical absorption spectra taken at increasing
temperatures up to 363 K provide detailed energy-temperature
trajectories for the exciton and trion absorption peaks, as shown
in Fig.~\ref{exp} (c). The absorption spectrum at 300K generally
agrees with previous experimental results \cite{2010wang,2010mak},
and ultimately reveals that both the exciton and trion peak
positions incur red shifts of roughly 50 meV relative to their 4 K
values. Corresponding PL spectra measurements indicate similar
frequency shifts, as shown in Fig.~\ref{exp} (d).

In order to understand these frequency shifts, we focus on the
impacts of lattice expansion for the following reasons: 1)
Previous studies show that the variation of the lattice constant
of monolayer MoS$_2$ will induce substantial changes in its
absorption spectra. \cite{2012liu} 2) The exciton and trion states
of monolayer MoS$_2$ are fundamentally tied to its band gap.
Accordingly, the energy-temperature trajectories plotted in
Figs.~\ref{exp} (c) and (d) show that these states display a
similar temperature dependence and thus their frequency shifts
likely share the same physical origin.

We thus present results from a first-principles simulation of
fully relaxed monolayer MoS$_2$, for which the in-plane lattice
constant is varied to mimic thermal expansion effects. We also
break all symmetries for more realistic simulations. The
quasiparticle band structure and optical absorption spectrum is
obtained for each structure for the sake of comparison with
experimental results. The simulation uses density functional
theory (DFT) and the single-shot G$_0$W$_0$ approach
\cite{1986Hybertsen}, which closely follows the approach taken in
Ref. \cite{2000Rohlfing}. A fine k-grid (60 x 60 x 1) in the first
Brillouin zone (BZ) is interpolated from a coarse grid (24 x 24 x
1) when finding converged excitonic states. The optical absorption
spectrum is then calculated by solving the BSE, incorporating four
valence bands, four conduction bands, and using incident light
polarized parallel to the monolayer sheet. Spurious interactions
between neighboring layers are avoided by imposing a slab Coulomb
truncation. Spin-orbital coupling is not considered in the
simulations.

\begin{figure}
\includegraphics*[scale=0.40]{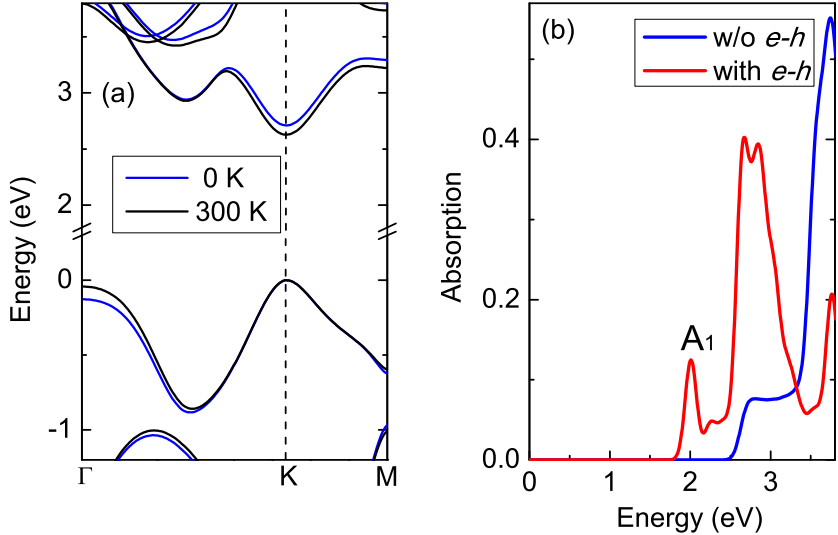}
\caption{(Color online) (a) The quasiparticle band structure of
monolayer MoS$_2$. The blue curves are about zero temperature
while the dark curves include the lattice expansion at 300 K. (b)
The optical absorption spectra of monolayer MoS$_2$ at zero
temperature with and without \emph{\emph{e-h}} interactions included,
respectively. A 0.06-eV Gaussian smearing is applied to obtain the
smoothed spectra.} \label{band}
\end{figure}

First, we focus on the zero-temperature limit structure, which is
fully relaxed according to both forces and stresses. The in-plane
lattice constant is 3.18 $\AA$. The quasiparticle-corrected band
structure of monolayer MoS$_2$ at 0 K is depicted in Fig.
~\ref{band} (a). The calculated structure contains a direct band
gap of size 2.63 eV. This gap is widened from its DFT value (1.69
eV) significantly by many-electron interactions because of
depressed screening. The optical absorption spectra of monolayer
MoS$_2$ at 0 K, with and without \emph{e-h} interactions, are
summarized in Fig. ~\ref{band} (b). This low-dimensional
semiconductor's spectrum is dramatically affected by \emph{e-h}
interactions, as is typical for such systems
\cite{2009yang,2010rubio,2013yang}. Including \emph{e-h}
interactions lowers the optical absorption edge from 2.63 eV to
2.01 eV, indicating a large \emph{e-h} binding energy of 0.62 eV.
The location of the first prominent peak is close to our
low-temperature measurement (1.94 eV) shown in Fig. ~\ref{exp}. It
is also close to recent calculations
\cite{exciton-1,exciton-2,exciton-3,exciton-4}; the slight
differences may be due to our denser course k-point grid, which
does not change the quasiparticle band gap significantly but more
affects the excitonic effects. Due to the similar reason, we
observe one prominent absorption peak (marked as A1 in Fig.
~\ref{band} (b)) around the absorption edge, indicating that the
experimentally observed double-peak feature \cite{2010mak} results
from spin-orbital coupling, which is not included in this
simulation.

\begin{figure}
\includegraphics*[scale=0.40]{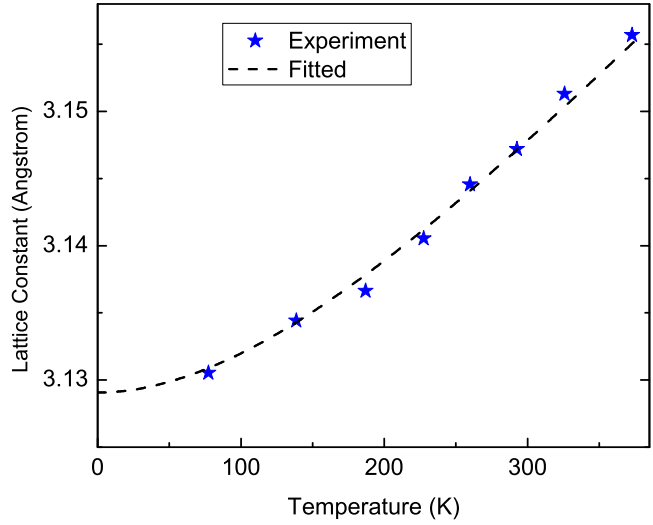}
\caption{(Color online) The in-plane lattice constant of bulk
MoS$_2$ variation according to the temperature. The blue stars are
experimental data \cite{1968young} and the dashed line is the
fitted curve, respectively.} \label{lattice}
\end{figure}

\begin{figure}
\includegraphics*[scale=0.50]{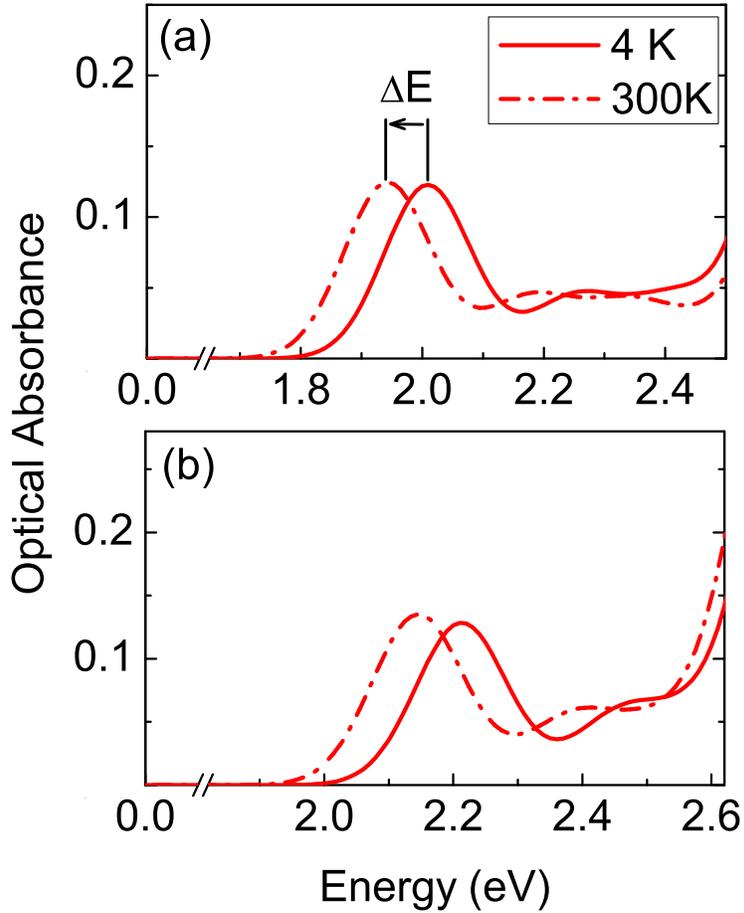}
\caption{(Color online) (a) The optical absorption spectra of
monolayer MoS$_2$ under different temperature when considering the
lattice thermal expansion effect. The zero-temperature lattice
constant is the theoretically optimized structure under DFT/PBE.
(b) The similar plots are in (a), but the zero-temperature lattice
constant is the experimental one read from Fig. 2. Excitonic
effects are included in all plots. A 60-meV Gaussian broadening is
applied.} \label{shift}
\end{figure}

Next, we consider effects due to lattice thermal expansion.
Presently, there is no available experimental data detailing the
expansion of monolayer MoS$_2$ as temperature is varied. It is
therefore natural to reference measured TEC values in multilayer
MoS$_2$ \cite{1968young} for our calculations. The presence of
only weak van der Waals (vdW) interactions between the bulk layers
suggests that the TEC of monolayer MoS$_2$ would not vary
drastically from that of multilayer MoS$_2$. Fig. ~\ref{lattice}
shows that the experimentally measured lattice constant grows from
roughly 3.13 $\AA$ to 3.15 $\AA$ when temperature is increased
from 0 K to 300 K, around a 0.6$\%$ expansion. Using this TEC, the
theoretical lattice constant for monolayer MoS$_2$ is increased
from 3.18 $\AA$ to 3.20 $\AA$ to mimic the increase in
temperature. The resulting quasiparticle band structure is
presented in Fig. ~\ref{band} (a). One can immediately note the
reduced band gap, which has shrunk from 2.63 eV to 2.55 eV as the
lattice expanded. This is qualitatively consistent with the red
shift seen in the experimental optical absorption spectra (Figs.
~\ref{exp} (a) and (c)). The calculated optical absorption
spectra, with \emph{e-h} interactions included, are depicted in
Fig.~\ref{shift} (a) for the effective 4 K and 300 K lattice
constants. The lattice expansion induces a 67-meV red shift of the
exciton peak from 2.01 eV to 1.95 eV. It is promising to see that
\emph{the effects of expanding the lattice in the calculations
satisfactorily agree with the experimentally observed temperature
effects, which are characterized by a 50 meV red shift}.

The presented measured and calculated results for monolayer
MoS$_2$ allow for a more realistic comparison of the
experimentally and theoretically derived optical absorption
spectra. In the low-temperature limit, the measured and
theoretical values for the exciton peak position are 1.94 eV and
2.01 eV, respectively. At room temperature, these respective
values are red shifted to 1.89 eV and 1.94 eV. Moreover, if the
spin-orbital splitting (around 140 meV \cite{exciton-3,exciton-4})
is included approximately, the optical absorption edge will be
lowered by about 70 meVs, locating the absorption peak at 1.94 eV.
This is exceptionally consistent with the low-temperature
measurement in Fig.~\ref{exp} (a). Although this perfect
consistence may not be conclusive because slight changes of
simulation setups can easily vary the result for a few tens meVs,
the present theoretical treatment of monolayer MoS$_2$, with the
thermal expansion included, shall be in even better agreement with
experiment, which is outstanding for a parameter-free calculation.

Additional calculations are performed using the lattice constants
for bulk MoS$_2$, as listed in Fig.~\ref{lattice}
\cite{1968young}, because it is known that DFT-optimized lattice
constant may be different from experimental measurements within
1$\%$. In the low-temperature limit, an extrapolated lattice
constant approximate to 3.13 $\AA$ is used. At room temperature,
we use the bulk value 3.15 $\AA$. The resulting optical absorption
spectra, with \emph{e-h} interactions included, is presented in
Fig. ~\ref{shift} (b). The exciton peak positions at 4 K and 300 K
are 2.20 eV and 2.13 eV, respectively, corresponding to a 65 meV
red shift with increasing temperature. Importantly, the frequency
shift of the exciton peak is thus nearly independent of the exact
lattice constants used. In this sense, tracking the frequency
shift of absorption peaks across temperatures serves as a
convenient method of estimating the TEC of monolayer MoS$_2$, and
perhaps in other materials as well, since it is extremely
challenging to directly measure the lattice constant of monolayer
materials at low temperature.

It is important to consider other factors that may have affected
the band gap and optical spectrum of MoS$_2$ in our experiments.
Higher-order \emph{e-ph} interactions are capable of renormalizing
band gaps in carbon semiconductors, however MoS$_2$ contains
relatively heavy constituents, thus it is expected that the
adiabatic approximation is valid. Additionally, band-filling
effects \cite{1998landin, 2008wang} are not expected to be
significant given the 2.6 eV-wide direct gap seen in
Fig,~\ref{band} (a), which diminishes thermal generation. Finally,
our samples are on a SiO$_2$ substrate, which may affect
many-electron effects by providing additional screening. According
to previous works \cite{1996neaton}, the variation of the
screening will affect both \emph{e-e} and \emph{e-h} interactions,
but in opposite directions. Therefore, the quasiparticle band gap
may vary but the final optical spectra and their shift would not
change significantly.

From the other point of view, the small difference of the shift of
optical peaks between experiments and simulations may reflect the
different TECs of monolayer and mutilayer MoS$_2$. Under the
first-order approximation, the optical gap is linearly
proportional to the lattice constant. Therefore, based on our
simulation result, the TEC of monolayer MoS$_2$ is around 75$\%$
of that of bulk counterpart. This trend is similar to the TECs of
graphene and graphite \cite{2005marzari,2009lau,2011son}.

In conclusion, comprehensive experimental measurements and
theoretical simulations have been performed to study the
temperature effects on the electronic structure and optical
excitations of monolayer MoS$_2$. Optical absorption and PL
measurements reveal that exciton and trion absorption peaks incur
redshifts as temperature is increased. First-principles
calculations reveal that these red shifts are mainly due to the
thermal expansion of the in-plane lattice constant. These results
provide a needed bridge between incongruent experiment and
simulation conditions that have generally been overlooked until
present. Additionally, it is demonstrated that tracking the shift
of absorption peak positions with temperature serves as a
convenient way of estimating the TEC of 2D MoS$_2$.
\begin{acknowledgement}

R.S., Y.L., and L.Y. are supported by NSF Grant No. DMR-1207141.
F. R. would like to acknowledge support from CCMR under NSF grant
number DMR-1120296, AFOSR-MURI under grant number
FA9550-09-1-0705, and ONR under grant number N00014-12-1-0072. The
computational resources have been provided by Lonestar of Teragrid
at the Texas Advanced Computing Center. The ground state
calculation is performed by the Quantum Espresso
\cite{2009Giannozzi}. The GW-BSE calculation is done with the
BerkeleyGW package \cite{2012Deslippe}.

\end{acknowledgement}




\end{document}